# The effective method to calculate eigenvalues of Chandrasekhar-Page angular equations

V.P.Neznamov[1,2*], I.I.Safronov[1]

[1]RFNC-VNNIEF, Russia, Sarov, Mira pr., 37, 607188
[2]National Research Nuclear University MEPhI, Moscow, Russia

Abstract

An effective, reliable and time saving numerical method with using of the Pruefer transformation is proposed to calculate eigenvalues of Chandrasekhar-Page angular equations.

Key words: angular equations, Pruefer transformation, spin-weighted mass-dependent spheroidal harmonics, eigenvalues, phase functions

[*] E-mail: neznamov@vniief.ru

# Introduction

In 1976, Chandrasekhar [1] showed how variables were separated in a Dirac massive equation in Kerr spacetime [2]. Page [3] extended this analysis to the Kerr-Newman spacetime [4]. In [1], [3], the wave function of the Dirac particle was represented as

$$\psi(t,r,\theta,\phi) = e^{-i\omega t + im\phi}\psi_{\omega m}(r,\theta). \tag{1}$$

The bispinor components $\psi_{\omega m}$ can be expressed as a product of two radial and two angular functions $\{R_+(r), R_-(r), S_+(\theta), S_-(\theta)\}$, obeying coupled ordinary differential equations.

Over the recent years, rather a large number of papers (see, for instance, [5] - [15]), are devoted to study of properties and methods of solving Chandrasekhar-Page angular equations. Among the references, one can single out [11], where the authors found numerical solutions to angular equations for rather a large range of parameter variations. The authors compared their results with those from the previous papers and indicated some of the errors in the literature.

In the current work, we have reproduced some eigenvalues of angular equations derived in [11] by using the Pruefer transformation [16], [17] for the original equations with appropriate boundary conditions specified after the transformation. This method turned out to be extremely effective, reliable and time saving.

In section 1, we introduce basic equations and denotations. Our denotations coincide with those from [11]. In section 2, we perform Pruefer transformation, discuss the method to solve the transformed equations. In Section 3 a part of calculated eigenvalues of Chandrasekhar-Page angular equations is presented and the obtained results are discussed.

## 1. Basic equations

The Chandrasekhar-Page angular equations have the form of:

$$\frac{dS_-}{d\theta} + \left(\frac{1}{2}\cot\theta + a\omega\sin\theta - m\csc\theta\right)S_- = (\lambda + a\mu\cos\theta)S_+,$$
$$\frac{dS_+}{d\theta} + \left(\frac{1}{2}\cot\theta - a\omega\sin\theta + m\csc\theta\right)S_+ = -(\lambda - a\mu\cos\theta)S_-. \tag{2}$$

Here, $a = \dfrac{J}{M}$ is the angular momentum parameter of Kerr and Kerr-Newman metrics; $\omega$ and $\mu$ are frequency and mass of Dirac particle; $S_-(\theta), S_+(\theta)$ are mass-dependent spheroidal harmonics of spin one-half; $\lambda$ are eigenvalues of the equation (2). The solutions (2) depend on two continuous parameters, $a\omega$ and $a\mu$. For the specified $a\omega, a\mu$, the eigenstates $\{S_+(\theta), S_-(\theta), \lambda\}$ can be identified by three discrete numbers: the angular momentum $j = \frac{1}{2}, \frac{3}{2}, ....$, the azimuthal component of the angular momentum $m = -j, -j+1, ..., j$ and the parity $P = \pm 1$.

Below, we will make identifications of

$$S_\pm = {}_{\pm 1/2}S_{j,m,P}^{(a\omega, a\mu)} \quad \text{and} \quad \lambda = \lambda_{j,m,P}^{(a\omega, a\mu)}. \tag{3}$$

The equations (2) permit a number of symmetries:

$$\lambda = \lambda_{j,m,P}^{(a\omega, a\mu)} = -\lambda_{j,-m,-P}^{(-a\omega, a\mu)} = -\lambda_{j,m,-P}^{(a\omega, -a\mu)} = \lambda_{j,-m,P}^{(-a\omega, -a\mu)}, \tag{4}$$

$${}_s S_{j,m,P}^{(a\omega, a\mu)} = (-1)^{s-\frac{1}{2}} {}_s S_{j,m,-P}^{(a\omega, -a\mu)}(\theta), \tag{5}$$

$$= P(-1)^{m-\frac{1}{2}} {}_{-s} S_{j,-m,-P}^{(-a\omega, a\mu)}(\theta), \tag{6}$$

$$= P(-1)^{j+m} {}_{-s} S_{j,m,P}^{(a\omega, a\mu)}(\pi - \theta). \tag{7}$$

The knowledge of spectrum $\lambda$ in the quadrant $a\omega > 0$, $a\mu > 0$ is sufficient to determine the full spectrum. As $\mu = 0$, the equations (2) are reduced to the equations for spin-weighted spheroidal harmonics. In the case of rotation absence $a = 0$, the equations (2) are reduced to the equations for spin-weighted spherical harmonics. The material of this section is described in more detail in [11].

## 2. The Pruefer transformation

Let

$$\begin{aligned} S_-(\theta) &= S(\theta) \sin \Phi(\theta), \\ S_+(\theta) &= S(\theta) \cos \Phi(\theta), \end{aligned} \tag{8}$$

where

$$\frac{S_-(\theta)}{S_+(\theta)} = \text{tg}\Phi(\theta), \tag{9}$$

$$S(\theta) = \left(S_-^2(\theta) + S_+^2(\theta)\right)^{1/2}. \tag{10}$$

Then, the equations (2) is written as

$$\frac{d\Phi}{d\theta} = \lambda + a\mu \cos\theta \cos(2\Phi) + \left(\frac{m}{\sin\theta} - a\omega \sin\theta\right)\sin(2\Phi), \tag{11}$$

$$\frac{d\ln S}{d\theta} = -\frac{1}{2}\operatorname{ctg}\theta + \left(a\omega\sin\theta - \frac{m}{\sin\theta}\right)\cos(2\Phi) + a\mu\cos\theta\sin(2\Phi). \tag{12}$$

Let us note that the solution (12) is determined only upon obtaining the solution (11). Below, we will deal with the problem of determining eigenvalues $\lambda$, i.e., with the solution of equation (11).

## 2.1 Boundary conditions

The asymptotics of the functions $S_-(\theta), S_+(\theta)$ in the vicinity of $\theta = 0$ and $\theta = \pi$ can be represented in the form of [12], [18]

$$S_-(x) = K(x) F_{1,2}(x), \tag{13}$$

$$F_1(x)\big|_{x\to 1} = \sum_{n=0}^{\infty} (F_1)_n (1-x)^n, \tag{14}$$

$$F_2(x)\big|_{x\to -1} = \sum_{n=0}^{\infty} (F_2)_n (1+x)^n, \tag{15}$$

$$K(x) = (1-x)^{\frac{1}{2}\left|m-\frac{1}{2}\right|} (1+x)^{\frac{1}{2}\left|m+\frac{1}{2}\right|}. \tag{16}$$

Here, $x = \cos\theta$.

The asymptotics $S_+(\theta)$ as $\theta \to 0$ or $\theta \to \pi$ can be easily determined from the relation of symmetry (7)

$$S_-(\theta) = P(-1)^{j+m} S_+(\pi - \theta). \tag{17}$$

Taking into account (9) and (13) - (17), the phase function $\Phi(\theta)$ as the poles $\theta = 0$ and $\theta = \pi$ is

$$\text{for } m < 0: \quad \Phi(0) = k\pi, \ \Phi(\pi) = \frac{\pi}{2} + k\pi, \ k = 0, \pm 1, \pm 2... \tag{18}$$

$$\text{for } m > 0: \quad \Phi(0) = \frac{\pi}{2} + k\pi, \ \Phi(\pi) = k\pi, \ k = 0, \pm 1, \pm 2... \tag{19}$$

As $\theta = \frac{\pi}{2}$, it follows from (17) that

$$S_-\left(\frac{\pi}{2}\right) = P(-1)^{j+m} S_+\left(\frac{\pi}{2}\right), \tag{20}$$

$$\operatorname{tg}\Phi\left(\frac{\pi}{2}\right) = P(-1)^{j+m} = \pm 1. \tag{21}$$

## 2.2 A numerical method to solve equations for the phase function $\Phi(\theta)$

The boundary conditions (18) - (21) show that in order to determine eigenvalues $\lambda$ one can use half the domain of functions $S_-(\theta), S_+(\theta)$: either $\theta \in \left[0, \frac{\pi}{2}\right]$ or $\theta \in \left[\frac{\pi}{2}, \pi\right]$. In the first case, by using the condition (21) during the numerical integration (11), one can follow in the inverse direction from $\theta = \frac{\pi}{2}$ to $\theta = 0$ and, when the condition (18) is met, obtain the set of eigenvalues $\lambda$. In the second case, the integration (11) should proceed from $\theta = \frac{\pi}{2}$ to $\theta = \pi$, and provided the condition (19) is met, the same spectrum $\lambda$ as that in the first case will be obtained.

The equation (11) has singularity $\frac{1}{\sin \theta}$ as $\theta \to 0$ and as $\theta \to \pi$. In calculations, the sufficient accuracy degree of determining a spectrum $\lambda$ is obtained with the restriction of the domain $\theta_{min} = 10^{-6} \div 10^{-9}$ and $\pi - \theta_{min} = \pi - \left(10^{-6} \div 10^{-9}\right)$.

To solve the equation (11) with the boundary conditions (18) - (21), we used the shooting method that includes the fifth-order Runge-Kutt implicit method with the step control (the Ehle scheme of Radau IIA three-stage method [19]) and the Bisection Method [20]. Below, we determined the eigenvalues $\lambda$ by means of the integration of (11) from $\theta = \frac{\pi}{2}$ to $\theta = \theta_{min}$.

## 3. Results

Let us first consider the solution (11) for a massless Dirac particle $(\mu = 0)$. This will be done for the parameters of

$$j = \frac{1}{2},\ l = 0,\ P = -1,\ m = +\frac{1}{2},\ a\omega = 1. \qquad (22)$$

In this case, according to (21), $\text{tg}\Phi\left(\frac{\pi}{2}\right) = 1$. In [11] (Appendix B, Table II), the value $\lambda$ calculated by the authors is

$$\lambda_{[11]} = -0,431544. \qquad (23)$$

Fig. 1 presents the function $\Phi(\lambda)\big|_{\theta = \theta_{min}}$ numerically determined by us for the values of the parameters (22). In the calculations, we considered the interval $\lambda \in [-10, 0], \theta_{min} = 10^{-9}$, the number of computational points for $\lambda$: $N_\lambda = 100$. The computational time on a commodity cjmputer is ~ 0.1s. The characteristic feature of the jump function $\Phi(\lambda)\big|_{\theta=\theta_{min}}$ is its variations by $\pi$ for eigenvalues $\lambda$ [17]. It is well seen in fig.1,2.

Since the equation (11) does not explicitly depend on $j$, each calculation with a specified value $m$ contains information on values $\lambda$ for all $j$ with the same multiplier $(-1)^{j+m}$. For instance, fig.1 provides information on values $\lambda$ as $m=+\frac{1}{2}$ and $j=\frac{1}{2},\frac{5}{2},\frac{9}{2}$, etc. If one changes the boundary condition $\text{tg}\Phi\left(\frac{\pi}{2}\right)=+1$ by $\text{tg}\Phi\left(\frac{\pi}{2}\right)=-1$, one can obtain the set of values $\lambda$ as $m=+\frac{1}{2}$ and $j=\frac{3}{2},\frac{7}{2},\frac{11}{2}$, etc. This set is also presented in fig.1.

If one shifts the initial interval $\lambda$ in calculations towards negative numbers bigger in modulus, we will obtain eigenvalues of $\lambda$ as $m=+\frac{1}{2}$ and arbitrary high $j$.

In the discussed computations, in addition to negative eigenvalues $\lambda$, which agree with the computations [11], as the parity $P=-1$, there formally exist jumps of the phase function as positive $\lambda$. We don't take into account such sets of values $\lambda$ in our work.

In the computation represented in fig.1, the value $\lambda$ as $j=\frac{1}{2}$ and $m=+\frac{1}{2}$ is

$$\lambda_1 = -0,431544. \qquad (24)$$

The value $\lambda$ in (24) coincides with $\lambda$ in (23) within the accuracy of six digits after a decimal point.

In [11] (Appendix B, Table VI) for the values of the parameters $j=\frac{3}{2}$, $l=1$, $P=-1$, $m=+\frac{1}{2}$, $a\omega=1$ for a massless Dirac particle the value $\lambda_{[11]}=-1,883249$ was calculated.

In our standard calculations with $\lambda\in[-10,0]$, $\theta_{min}=10^{-9}$, $N_\lambda=100$, $\text{tg}\Phi\left(\frac{\pi}{2}\right)=-1$ the value of $\lambda_2=-1,883249$ was also obtained (see fig. 1).

Now, let us consider the solution (11) for a massive Dirac particle $(\mu\neq 0)$ by using the following parameter values as an example

$$j=\frac{1}{2},\ l=0,\ P=+1,\ m=-\frac{1}{2},\ a\omega=1,\ \frac{\mu}{\omega}=0,4. \qquad (25)$$

In this case, according to (21), $\text{tg}\Phi\left(\frac{\pi}{2}\right)=1$. In [11] (Appendix B, Table I), the value $\lambda$ calculated for this case is

$$\lambda_{[11]} = 1,6425. \qquad (26)$$

In fig.2, the computed function $\Phi(\lambda)\big|_{\theta=\theta_{min}}$ is given for the values of the parameters (25). In the computation, there was considered the interval $\lambda \in [0, 10]$, $\theta_{min} = 10^{-9}$, $N_\lambda = 100$. The PC computational time is ~ 0.1s.

The obtained value $\lambda$ as $j=\frac{1}{2}$, $m=-\frac{1}{2}$ is

$$\lambda_1 = 1,6425, \qquad (27)$$

which coincides with the value $\lambda$ in (26), calculated by the authors of [11].

Fig.2 also presents information on values $\lambda$ as $m=-\frac{1}{2}$ and $j=\frac{1}{2},\frac{5}{2},\frac{9}{2}$ etc. When substituting the boundary condition $\text{tg}\Phi\left(\frac{\pi}{2}\right)=+1$ by $\text{tg}\Phi\left(\frac{\pi}{2}\right)=-1$, we obtain the set of values $\lambda$ as $m=-\frac{1}{2}$ and $j=\frac{3}{2},\frac{7}{2},\frac{11}{2}$, etc. (see fig.2)

In [11] (Appendix B, Table V), for a massive Dirac particle with the values of the parameters $j=\frac{3}{2}$, $l=2$, $P=+1$, $m=-\frac{1}{2}$, $a\omega=1$, $\frac{\mu}{\omega}=0,4$, the value of $\lambda_{[11]} = 2,343692$ was calculated.

In our computations with $\lambda \in [0, 10]$, $\theta_{min} = 10^{-9}$, $N_\lambda = 100$, $\text{tg}\Phi\left(\frac{\pi}{2}\right)=-1$, the value of $\lambda_2 = 2,343692$ was obtained (see fig. 2) that coincides with $\lambda_{[11]}$ with all digits after the decimal point.

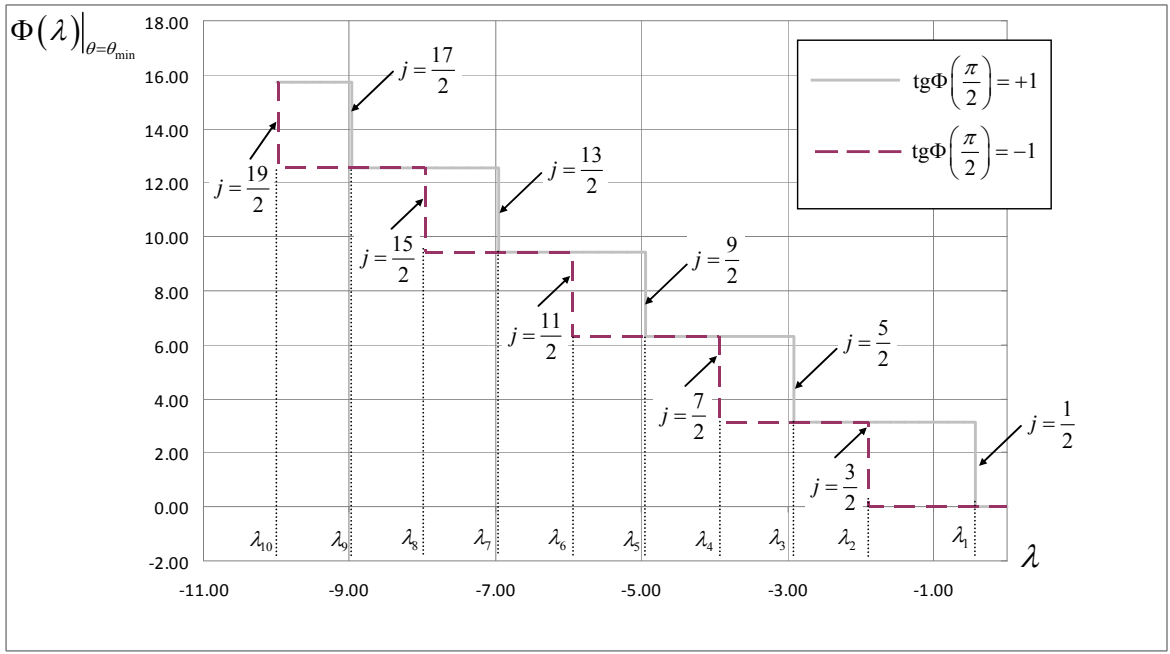

Figure 1. Dependence of the phase functions on a separation parameter $\lambda$. The eigenvalues $\lambda$ for different $j$ as $a\omega = 1,\ \mu = 0;\ P = -1,\ m = +1/2$.

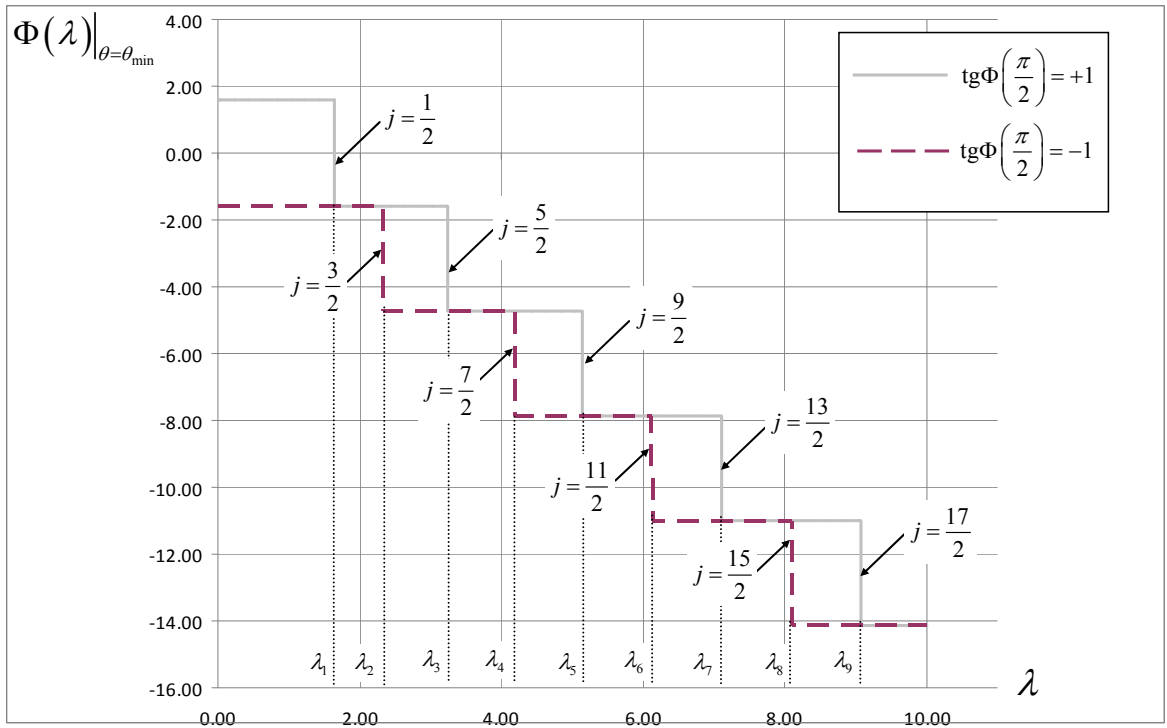

Figure 2. Dependence of the phase functions on a separation parameter $\lambda$. The eigenvalues $\lambda$ for different $j$ as $a\omega = 1,\ \mu/\omega = 0,4;\ P = +1,\ m = -1/2$.

Let us note that in the computations, while determining $\lambda$ with a high accuracy degree, the situation can arise when a jump $\Phi(\lambda)\big|_{\theta=\theta_{\min}}$ in the vicinity of eigenvalues $\lambda_k$ will be lower than $\pi$. To reconstruct the required jump with the specified accuracy degree, the appropriate decrease of $\theta_{min}$ is needed.

To summarize, we can draw the conclusion that the Pruefer transformation is an effective tool for time saving and reliable computations of eigenvalues of Chandrasekhar-Page angular equations.

## Acknowledgements

The authors would like to thank A.L. Novoselova for the essential technical support while elaborating the paper.

# References

[1] S. Chandrasekhar, Proc. R. Soc. London A 349, 571 (1976).

[2] R.P.Kerr, Phys. Rev.Lett. 11, 237 (1963).

[3] D. N. Page, Phys. Rev. D 14, 1509 (1976).

[4] E.T.Newman, E.Couch, K.Chinnapared, A.Exton, A.Prakash and R.Torrence, J.Math.Phys. 6, 918 (1965).

[5] E. G. Kalnins and W. Miller Jr., J. Math. Phys. 33, 286 (1992).

[6] K. G. Suffern, E. D. Fackerell, and C. M. Cosgrove, J. Math. Phys. 24, 1350 (1983).

[7] D. Batic, H. Schmid, and M. Winklmeier, J. Math. Phys. 46, 012504 (2005), [math-ph/0402047].

[8] S. K. Chakrabarti, Proc. R. Soc. London A 391, 27 (1984).

[9] S. Chandrasekhar, The Mathematical Theory of Black Holes (Oxford University Press, 1983).

[10] M. Winklmeier, Journal of Differential Equations 245, 2145 (2008), [arXiv:0806.1866].

[11] S.Dolan, J.Gair, Class. Quantum Grav. 26, 175020 (2009).

[12] E.W.Leaver, Proc. R. Soc. Lond A, 402, 285-298 (1985).

[13] M.K.-H. Kiessling and A.S.Tahvildar-Zadeh, J. Math. Phys. 56, 042303 (2015).

[14] Shahar Hod, arxiv: 1506.04148 [gr-qc].

[15] D.Batic, K.Morgan, M.Nowakowski and S.Bravo Medina, arxiv:1509.00452 [gr-qc].

[16] H.Pruefer, Math. Ann. 95, 499 (1926).

[17] I.Ulehla, J.Horejsi, Phys. Lett, 113A, №7, 355 (1986).

[18] C.L.Pekeris, K.Frankowski, Phys. Rev. A 39, 518 (1989).

[19] E.Hairer, G Wanner. Solving ordinary differential equations II. Stiff and Differential-Algebraic Problems, Second Revised Edition, Springer-Verlag 1991, 1996 (Russian translation – M.: Mir, 1999).

[20] W.H. Press, S.A. Teukolsky, W.T. Vetterling, B.P. Flannery. *Numerical Recipes in Fortran 77: The Art of Scientific Computing*, Second Edition, Cambridge University Press, Cambridge, UK, 1997.